# Short-Time Dynamics of Fe$_2$/V$_{13}$ Magnetic Superlattice Models

A. K. Murtazaev [a,b] and V. A. Mutailamov [a,*]

[a]Amirkhanov Institute of Physics, Daghestan Scientific Center, Russian Academy of Sciences, Makhachkala, 367003 Daghestan, Russia

[b]Daghestan State University, Makhachkala, 367025 Daghestan, Russia

*e-mail: vadim.mut@mail.ru

ABSTRACT

Critical relaxation from a low_temperature fully ordered state of Fe2/V13 iron–vanadium magnetic superlattice models has been studied using the method of short_time dynamics. Systems with three variants of the ratio $R$ of inter- to intralayer exchange coupling have been considered. Particles with $N = 262144$ spins have been simulated with periodic boundary conditions. Calculations have been performed using the standard Metropolis algorithm of the Monte Carlo method. The static critical exponents of magnetization and correlation radius, as well as the dynamic critical exponent, have been calculated for three $R$ values. It is established that a small decrease in the exchange ratio (from $R = 1.0$ to $0.8$) does not significantly influence the character of the short_time dynamics in the models studied. A further significant decrease in this ratio (to $R = 0.01$), for which a transition from three_dimensional to quasi_two_dimensional magnetism is possible, leads to significant changes in the dynamic behavior of iron–vanadium magnetic superlattice models.

## 1. INTRODUCTION

Considerable interest in the modern condensed state physics is devoted to investigating the properties of magnetic superlattices comprising alternating atomic layers of magnetic and nonmagnetic materials. Since the basic properties of these superlattices, such as magnetization, interlayer exchange coupling, and magnetoresistance, can be controlled by means of external factors, it is possible to create structures with preset parameters, which makes these materials unique objects for both practical applications and theoretical investigation [1–3]. In addition, magnetic superlattices provide an ideal opportunity to observe the continuous crossover from three_dimensional (3D) to two-dimensional (2D) magnetism and vice versa.

At present, the situation with investigating the critical properties of magnetic superlattices is rather ambiguous, because the available results in this field are contradictory [4, 5]. Experimental investigations of these systems require materials of very high quality. The creation of high-quality samples and high-precision characterization of their critical properties are extremely difficult tasks. This probably accounts both for the small number of works in which the magnetic superlattices were studied near the point of phase transition and for the contradictory results.

For this reason, in recent years, the critical properties of magnetic superlattices have also been studied by computational physics methods. In particular, we have studied [6–8] the static critical behavior of iron–vanadium (Fe/V) magnetic superlattice models, determined their static critical exponents, and elucidated their dependence on the ratio $R$ of inter- to intralayer exchange coupling. The such computational physics methods as Monte Carlo and molecular dynamics simulations possess a number of advantages related to both their strict mathematical justification and the possibility of controlling the error in the calculations proper, as well as to the possibility of estimating the influence of various parameters on the results of simulations.

It is also very interesting to consider the dynamic critical behavior of magnetic superlattices, which have remained almost unstudied so far.

Investigation of the dynamic critical properties of systems is among topical problems of modern statistical physics and the physics of phase transitions [9–11]. Significant achievements in this field have mostly been gained in theoretical and experimental investigations. Nevertheless, construction of a rigorous and consistent theory of dynamic critical phenomena based on microscopic Hamiltonians is among the central problems in the modern theory of phase transitions and critical phenomena, which is still far from being solved [9, 12].

Rigorous theoretical investigation of the critical dynamics of spin systems based on microscopic Hamiltonians is an extremely difficult task even for simple spin models. The situation with studying the critical properties of magnetic superlattice models is even more complicated. In particular, investigations of the static critical properties of iron–vanadium magnetic superlattices showed that their critical exponents depend on the interlayer exchange coupling parameters [6–8]. At the same time, scaling relations between the critical exponents remain valid to within high precision. This situation does not fall in the modern theory of phase transitions and critical phenomena. Therefore, studying the critical dynamics is not only of considerable independent interest, but it may be a key point in explaining difficulties encountered in investigating static critical phenomena.

In recent decades, the critical dynamics of magnetic material models have been successfully studied using the method of short_time dynamics [13–16]. For this, the critical relaxation of a magnetic model from a nonequlibrium to equilibrium state is studied using model A according to the Hohenberg–Halperin hierarchy of universality classes in the dynamic critical behavior [17]. It is traditionally believed that universal scaling behavior exists only in a state of thermodynamic equilibrium. However, it has been shown [18] that some dynamic systems can exhibit universal scaling behavior at early stages of their temporal evolution from a high_temperature disordered state to a state corresponding to the phase transition temperature. This behavior is achieved upon a certain period of time, which is sufficiently long on a microscopic scale but still short on a macroscopic scale. An analogous pattern is observed for a system evolving from a lowtemperature ordered state [14, 15].

## 2. METHOD OF INVESTIGATION

Using the renormalization group method, Janssen et al. [18] showed that, far from the point of equilibrium, the $k$th magnetization moment $M^{(k)}$ for a microscopically time short interval obeys the following scaling form:

$$M^{(k)}(t,\tau,L,m_0) = b^{-k\beta/\nu} M^{(k)}(b^{-z}t, b^{1/\nu}\tau, b^{-1}L, b^{x_0}m_0), \tag{1}$$

where $t$ is the current time; $\tau$ is the reduced temperature; $L$ is the linear size of the system; $b$ is the scaling factor; $\beta$ and $\nu$ are the static critical exponents of magnetization and correlation radius, respectively; $z$ is the dynamic critical exponent; and $x_0$ is a new independent critical exponent that determines the scaling dimension of initial magnetization $m_0$.

Starting from a low_temperature ordered state ($m_0 = 1$) at the critical point ($\tau = 0$) for a system with scaling factor $b = t^{1/z}$ in Eq. (1) and sufficiently large linear size $L$, the theory predicts a power behavior of the magnetization in a short-time regime:

$$M(t) \sim t^{-c_1}, \quad c_1 = \frac{\beta}{\nu z}. \tag{2}$$

Taking the logarithm on both sides of Eq. (2) and the derivative with respect to $\tau$ at $\tau = 0$, we obtain thefollowing power law for the logarithmic derivative of magnetization:

$$\partial_\tau \ln M(t,\tau)\big|_{\tau=0} \sim t^{-c_{l1}}, \quad c_{l1} = \frac{1}{\nu z}. \tag{3}$$

For the Binder cummulant $U_L(t)$ calculated using the first and second magnetization moments, the theory of finite-dimension scaling predicts the following temporal dependence at $\tau = 0$:

$$U_L(t) = \frac{M^{(2)}}{(M)^2} - 1 \sim t^{c_U}, \quad c_U = \frac{d}{z}. \tag{4}$$

Thus, using the short_time dynamics method in a single numerical simulation with Eqs. (2)–(4), it is possible to determine the values of three critical exponents: $\beta$, $\nu$, and $z$. In addition, by constructing temporal dependences (2) for various temperatures on a double logarithmic scale, it is possible to estimate the critical temperature $T_c$ by determining their deviation from linearity.

### 3. DESCRIPTION OF MODEL

According to the microscopic model of a $Fe_2/V_{13}$ iron–vanadium magnetic superlattice proposed earlier [6–8], each Fe atom has four nearest neighbors in the adjacent iron layer. Iron layers are shifted relative to each other by the lattice half-period in the $x$ and $y$ directions. The magnetic moments of iron atoms are ordered in the $xy$ plane. Figure 1 shows a schematic diagram of the iron–vanadium magnetic superlattice under consideration.

The interaction between nearest neighbors in each iron layer has a ferromagnetic character and is determined by the intralayer exchange coupling parameter $J_\parallel$. The interlayer coupling (characterized by $J_\perp$) is mediated by conduction electrons transferred via nonmagnetic vanadium spacer (RKKY interaction). In real superlattices, the magnitude and sign of this interaction can vary depending on the distance between iron layers, which depends in turn on the amount of hydrogen adsorbed on vanadium [4, 5]. Since the exact law of the RKKY interaction is unknown, it is conventional practice in numerical simulations to study the entire range of interlayer exchange coupling from $J_\perp = -J_\parallel$ to $J_\perp = J_\parallel$.

Since the spacing between magnetic layers in experiment is significantly greater than the interatomic distance, each Fe atom interacts with an averaged moment of adjacent layers. The size of the region of averaging is a parameter of the model. Our simulations have been carried out for a limiting case in which each atom interacts with a single nearest neighbor in the adjacent layer. Investigations [6–8] of the static critical behavior of magnetic superlattices showed that this approach gives the best description of critical behavior for the adopted models.

Thus, the Hamiltonian of the system under consideration can be written in the form of a modified 3D $XY$ model as follows:

$$H = -J_\parallel \frac{1}{2} \sum_{i,j} \left( S_i^x S_j^x + S_i^y S_j^y \right) - J_\perp \frac{1}{2} \sum_{i,k} \left( S_i^x S_k^x + S_i^y S_k^y \right), \qquad (5)$$

where the first sum describes the direct exchange coupling between each magnetic atom and its nearest neighbors inside the layer; the second sum accounts for the RKKY interaction with atoms of adjacent layers via the nonmagnetic spacer; and are projections of the spin localized at the $i$th lattice site. In our numerical experiments, the ratio $R = J_\perp/J_\parallel$ is a preset parameter that can vary from $R = -1.0$ to $R = 1.0$ [6–8].

According to the results of our investigations [6–8] of the static critical behavior of the $Fe_2/V_{13}$ ironvanadium magnetic superlattice, a gradual decrease in $R$ below unity leads to smooth variation of the critical exponents. Until a certain threshold is reached, the system obeys the well-known scaling relations (e.g., the Rushbrooke inequality [19]) between critical exponents. However, at $R = 0.01$, the critical exponents exhibit a significant change, which is accompanied by violation of the scaling relations. This behavior suggests that $R = 0.01$ is the boundary of the transition from 3D to quasi-2D magnetism.

In order to evaluate the influence of interlayer exchange coupling on the character of short-time dynamics in iron–vanadium magnetic superlattice models, we have selected three values of the ratio of inter- to intralayer exchange coupling: $R = 1.0$, $R = 0.8$, and $R = 0.01$. Note that, in the case of $R = 1.0$, the Hamiltonian of the model under consideration according to Eq. (5) is analogous to the Hamiltonian of the classical three-dimensional *XY* model.

## 4. RESULTS AND DISCUSSION

Using the method of short-time dynamics, we have studied the $Fe_2/V_{13}$ iron–vanadium magnetic superlattice model with a linear size of $L = 64$. Calculations have been performed using the standard Metropolis algorithm of the Monte Carlo method for particles containing $N = 262\ 144$ spins with periodic boundary conditions. The system exhibited relaxation from an initial low-temperature fully ordered state with $m_0 = 1$ for a period of time $t_{max} = 1\ 000$, where the temporal unit corresponds to one Monte Carlo simulation step per spin. The relaxation curves were calculated for up to 14 000 runs and the results were averaged.

The critical temperatures were determined from the temporal dependences of magnetization (2), which must be a straight line on a double logarithmic scale at the point of the phase transition. A deviation from linearity was determined by the least squares method. The critical temperature ($T_c$) was defined as that for which the deviation from linearity was at minimum. In determining $T_c$, the magnetization curves were analyzed at a step of $\Delta T = 0.01$ in units of the exchange integral $k_B T/J_\parallel$.

Figure 2 shows the temporal variation of magnetization at three temperatures for $R = 1.0$ in the vicinity of the phase transition point as plotted on a double logarithmic scale (here and below, all quantities are expressed in dimensionless units). The critical temperatures (in units of the exchange integral $k_B T/J_\parallel$) determined for the three $R$ values are presented in the table. The logarithmic derivative at the point of the phase transition was approximately calculated by the least squares using three temporal dependences of magnetization constructed for $T_c - 0.01$, $T_c$, and $T_c + 0.01$.

Figure 3 presents a double logarithmic plot of the Binder cummulant $U_L$ versus time $t$ at the phase transition point for three values of the exchange coupling ratio $R$. An analysis of these data showed that the power scaling relation for $U_L(t)$ is valid beginning with a time on the order of $t = 100$. For this reason, the final approximation by the least squares using formula (4) was carried out on the interval $t = [100; 1000]$. The corresponding values of exponents $C_U$ and $z$ for various $R$ are presented in the table.

As can be seen from Fig. 3, the $C_U$ curves for the exchange coupling ratios $R = 1.0$ and $0.8$ virtually coincide with each other and significantly differ from the curve for $R = 0.01$. An analogous behavior is observed for the values of exponents $C_U$ and $z$. The values of $z$ at $R = 1.0$ and $0.8$ are close to the theoretically predicted value of the dynamic critical exponent for anisotropic magnets ($z = 2$, model A [17]).

Figures 4 and 5 show the temporal variation of the magnetization and its derivative with respect to time, respectively, plotted on the double logarithmic scale at the phase transition point for three values of the exchange coupling ratio $R$. Upon approximation by least squares on the interval $t = [100; 1000]$, exponents $cl$ and $cl_1$ were calculated using formulas (2) and (3) and then used to determine the static critical exponents of magnetization ($\beta$) and correlation radius ($\nu$) from Eqs. (2)–(4). The results are summarized in the table. Similar to the case of the binder cummulant, the values of critical exponents for $R = 1.0$ and $0.8$ virtually coincide and significantly differ from the values for $R = 0.01$.

It should be noted that the values of critical exponents obtained in this study for the exchange coupling ratio $R = 1$ are close to the theoretically predicted values of $\beta = 0.3485(3)$ and $\nu = 0.67155(37)$ according to the classical three-dimensional *XY* model [20].

Thus, the results of our investigation showed that a small decrease in the exchange ratio (from $R = 1.0$ to $0.8$) does not significantly influence the character of the short_time dynamics in the $Fe_2/V_{13}$ iron–vanadium magnetic superlattice models studied. A significant decrease in this ratio (to $R = 0.01$), for which a transition from three_dimensional to quasi-twodimensional magnetism is possible according to [6–8], leads to significant changes in the dynamic behavior of models. which is accompanied by significant changes in the values of critical exponents of magnetization ($\beta$) and correlation radius ($\nu$) and the dynamic critical exponent ($z$).

ACKNOWLEDGMENTS

This study was supported in part by the Ministry of Education and Science of the Russian Federation (project no. 14.B37.21.1092 "Development and Investigation of Promising Nanostructure Models by Computer Simulation Methods") and the Russian Foundation for Basic Research (project nos. 10_02_00130 and 12_02_96504_r_yug_a).

*Translated by P. Pozdeev*

Table 1. Critical exponents and critical temperatures of iron–vanadium magnetic superlattices with various $R$ values

| $R$ | 1.0 | 0.8 | 0.01 |
|---|---|---|---|
| $k_b T_c / J_\parallel$ | 1.752(1) | 1.669(1) | 1.044(1) |
| $C_I$ | 0.29(3) | 0.27(3) | 0.13(3) |
| $C_{lI}$ | 0.79(3) | 0.78(3) | 0.53(3) |
| $C_U$ | 1.54(3) | 1.51(3) | 1.20(3) |
| $\beta$ | 0.36(3) | 0.35(3) | 0.25(3) |
| $\nu$ | 0.65(3) | 0.65(3) | 0.75(3) |
| $z$ | 1.95(3) | 1.99(3) | 2.51(3) |

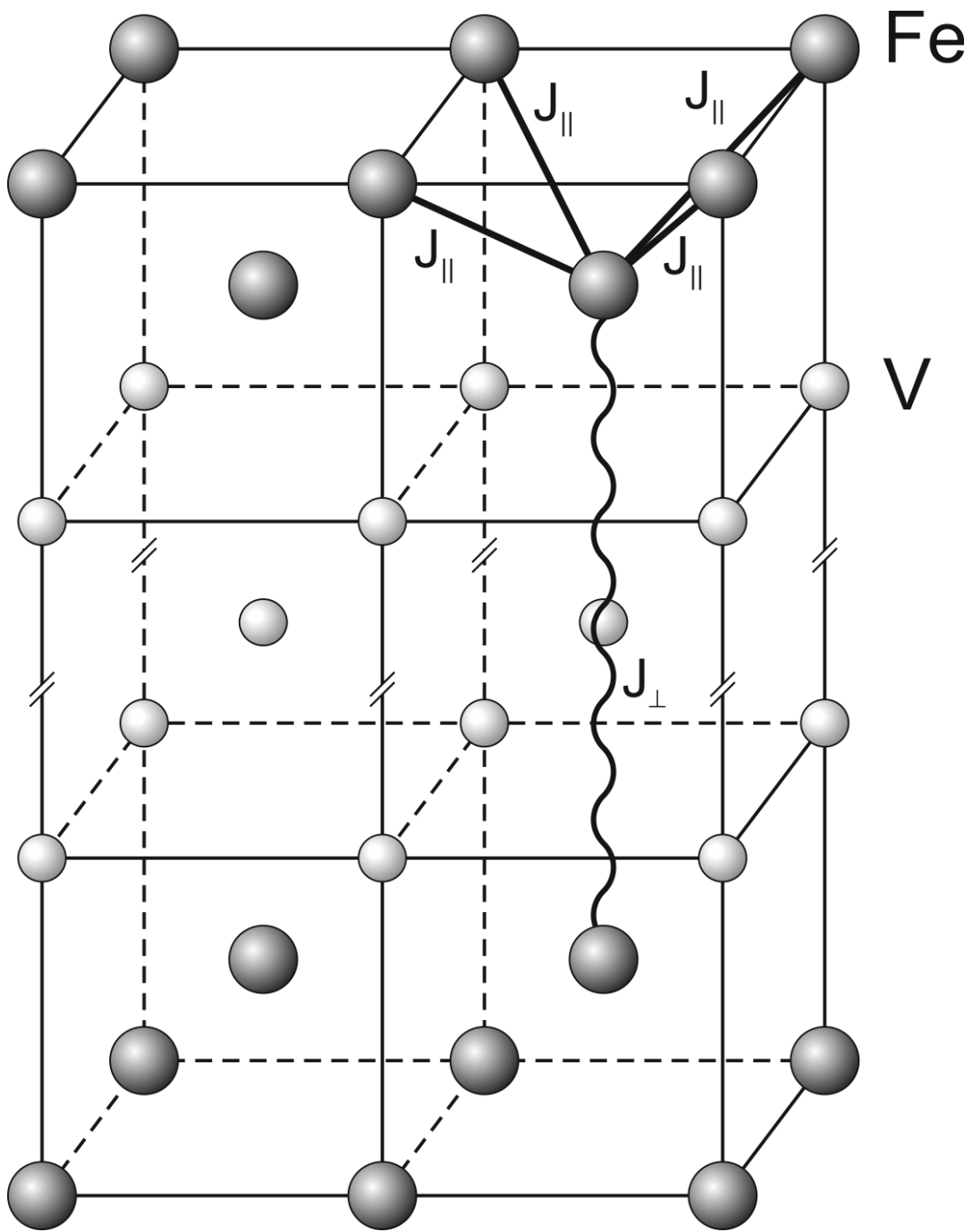

Fig. 1. Schematic diagram of Fe2/V13 iron–vanadium magnetic superlattice. For convenient illustration, only three (of the total 13) monolayers of vanadium are depicted.

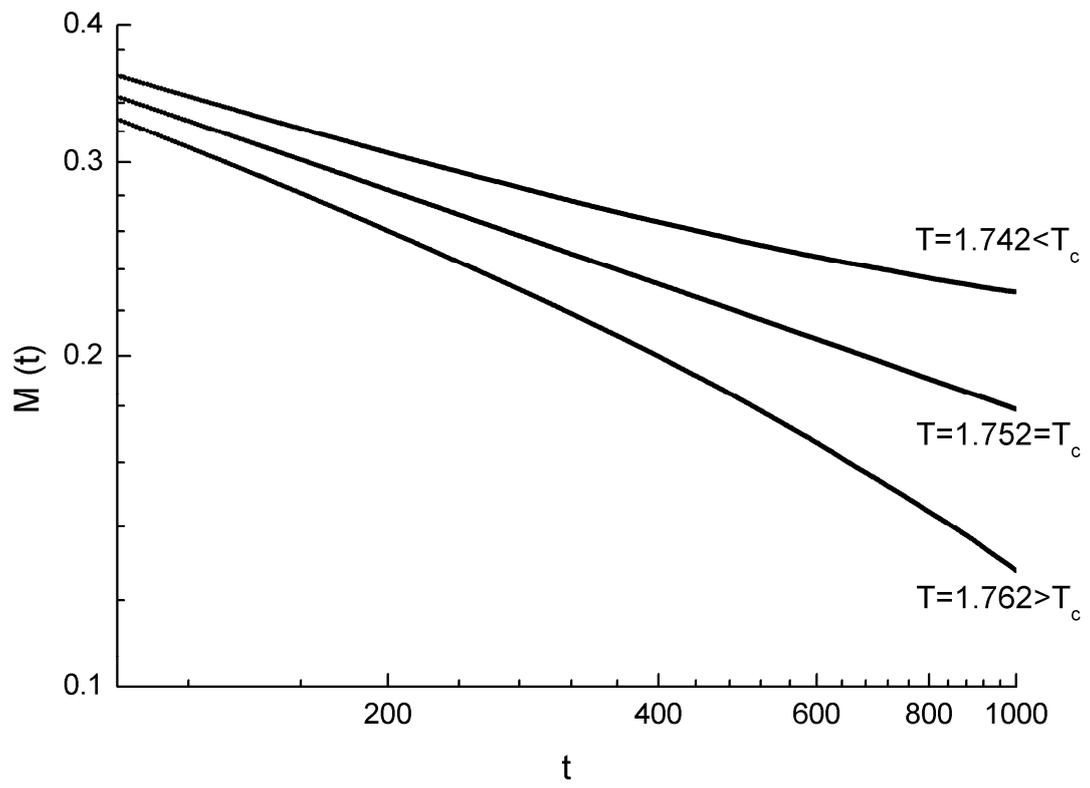

Fig. 2. Temporal variation of magnetization at various temperatures for *R* = 1.0.

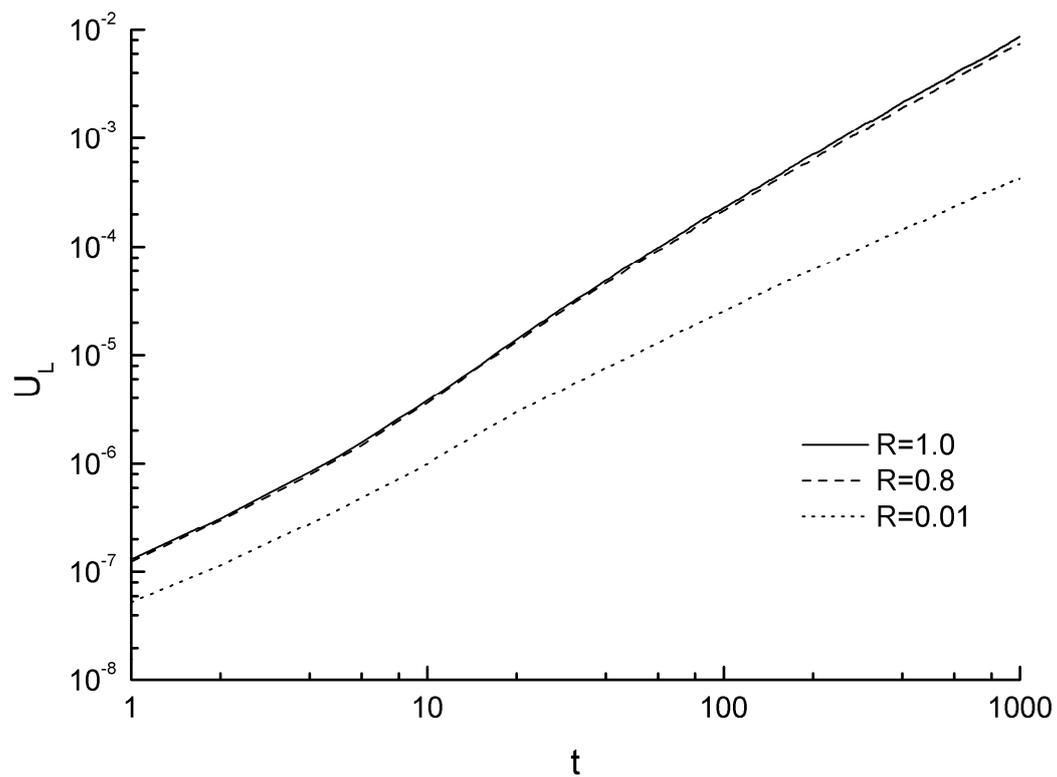

Fig. 3. Double logarithmic plots of Binder cummulant $U_L$ versus time $t$ at phase transition point for three values of exchange coupling ratio $R$.

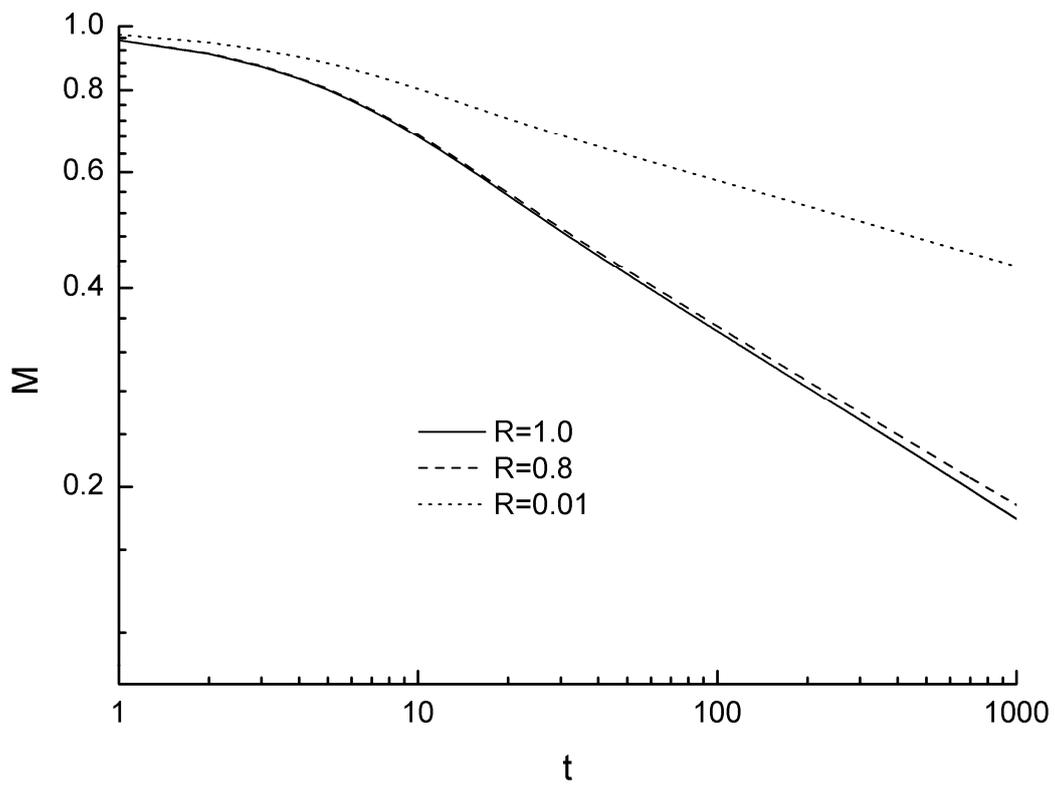

Fig. 4. Temporal variation of magnetization at phase transition point for three values of exchange coupling ratio *R*.

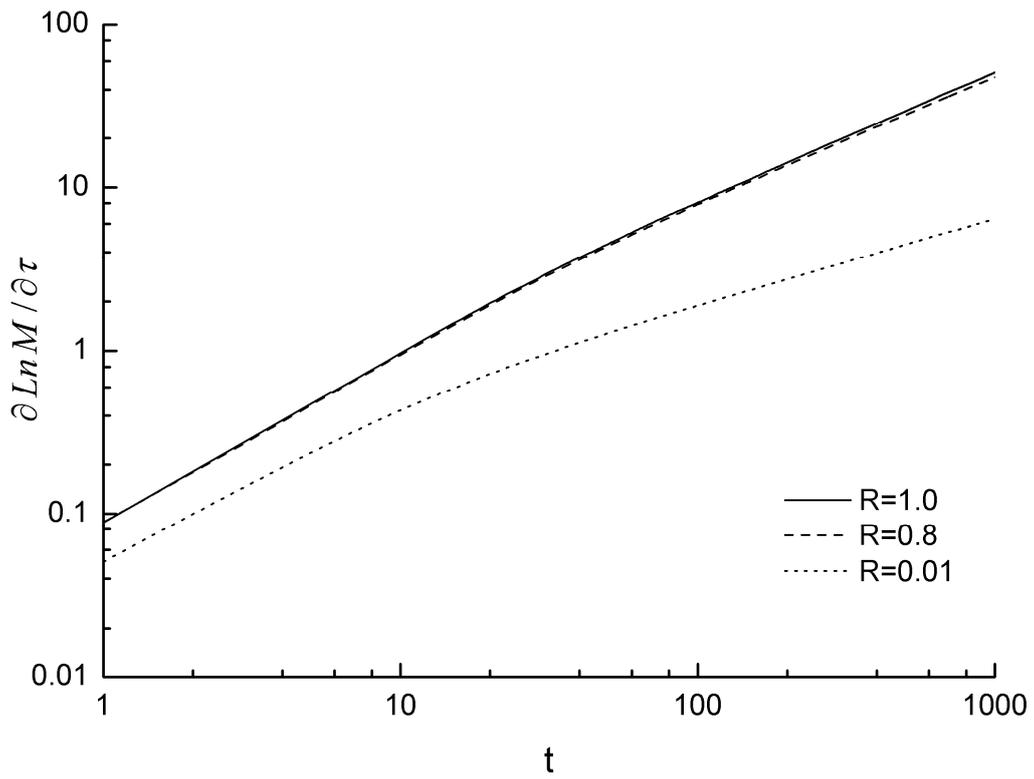

Fig. 5. Temporal variation of logarithmic derivative of magnetization at phase transition point for three values of exchange coupling ratio *R*.